\documentstyle[prl,aps] {revtex}                     
\begin{document}
\title{Measurement of the recoil polarization in the
       $\boldmath p(\vec e, e' \vec p\,)\pi^{0}$  reaction at the 
       $\Delta(1232)$  resonance }

\author{
Th.~Pospischil$^{1}$\thanks{comprises part of doctoral thesis}, 
P.~Bartsch$^1$, 
D.~Baumann$^1$,  
J. Bermuth$^2$,
R.~B\"ohm$^1$, 
K.~Bohinc$^{1,3}$, 
S.~Derber$^1$, 
M.~Ding$^1$, 
M. Distler$^1$, 
D.~Drechsel$^1$, 
D. Elsner$^1$, 
I.~Ewald$^1$, 
J.~Friedrich$^1$, 
J.M.~Friedrich$^1$\thanks{present address: Physik Department E18, 
                                           TU M{\"u}nchen, Germany}, 
R.~Geiges$^1$,
S.~Hedicke$^1$,
P.~Jennewein$^1$, 
M.~Kahrau$^1$,
S.S.~Kamalov$^1$\thanks{permanent address: Laboratory for Theoretical Physics,
                                           JINR Dubna, Russia}, 
F.~Klein$^1$, 
K.W.~Krygier$^1$, 
J.Lac$^4$,
A.~Liesenfeld$^1$, 
J.~McIntyre$^4$, 
H.~Merkel$^1$, 
P.~Merle$^1$,
U.~M\"uller$^1$, 
R.~Neuhausen$^1$, 
M.~Potokar$^3$, 
R.D.~Ransome$^4$,
D. Rohe$^{2,5}$, 
G.~Rosner$^1$\thanks{present address: Dept. of Physics and Astronomy,
                                      University of Glasgow, UK}, 
H.~Schmieden$^1$\thanks{corresponding author, email: hs@kph.uni-mainz.de}, 
M.~Seimetz$^1$, 
S.~\v Sirca$^3$\thanks{present address: Laboratory for Nuclear Science,
                                        MIT, Cambridge, MA, USA}, 
I.~Sick$^5$, 
A.~S\"ule$^1$, 
L.Tiator$^1$,
A.~Wagner$^1$, 
Th.~Walcher$^1$, 
G.A.~Warren$^5$,
and M. Weis$^1$ 
       }
\address{$^1$  Institut f\"ur Kernphysik, Universit\"at Mainz, D-55099 Mainz,  
               Germany \\                                                    
         $^2$  Institut f\"ur Physik, Universit\"at Mainz, D-55099 Mainz, 
               Germany \\                   
         $^3$  Institut Jo\v zef Stefan, University of Ljubljana,            
               SI-1001 Ljubljana, Slovenia \\                                
         $^4$  Rutgers University, Piscataway, NJ, USA \\
         $^5$  Dept. f\"ur Physik und Astronomie, Universit\"at Basel,       
               CH-4056 Basel, Switzerland \\                                 
        }                                                                    

\date{\today}

\maketitle

\begin{abstract}
The recoil proton polarization has been measured in the 
$ p(\vec e,e'\vec p\,)\pi^0 $  reaction in parallel kinematics
around $ W = 1232 $\,MeV, $ Q^2 = 0.121 $\,(GeV/c)$^2$  and 
$\epsilon = 0.718$  using the polarized c.w. electron
beam of the Mainz Microtron.
Due to the spin precession in a magnetic spectrometer, all three
proton polarization components 
$P_x/P_e = (-11.4 \pm 1.3 \pm 1.4)$\,\%,
$P_y     = (-43.1 \pm 1.3 \pm 2.2)$\,\%, and 
$P_z/P_e = ( 56.2 \pm 1.5 \pm 2.6)$\,\%
could be measured simultaneously.
The Coulomb quadrupole to magnetic dipole ratio
$\text{CMR} = (-6.4\pm 0.7_{stat}\pm 0.8_{syst})$\,\%
was determined from $P_x$ in the framework of the Mainz Unitary Isobar Model.
The consistency among the reduced polarizations and the extraction of
the ratio of longitudinal to transverse response is discussed.

\end{abstract}
\twocolumn
%
%

The interaction between the three constituent quarks in a nucleon is
mediated by gluons and, at long range, pions.
Both exchange bosons produce a tensor force and, consequently,
the nucleon wave function 
might have a D-state admixture \cite{deRujula75},
as is the case for the deuteron.
Such an admixture results in an intrinsic deformation of the nucleon
\cite{Glashow79}.
However, the spectroscopic quadrupole moment of the nucleon must vanish due 
to its spin 1/2.
In the case of the first excited state $\Delta(1232)$  with spin 3/2
the short life-time prohibits a direct measurement of the quadrupole moment.
On the other hand, in the electromagnetic $N \rightarrow \Delta(1232)$  
excitation a D-admixture will be visible as small quadrupole admixtures
to the dominating M1 transition.
In the decay of the $\Delta(1232)$  resonance into the 
$N \pi$  channel, this quadrupole mixing is associated with non-zero 
electric quadrupole to magnetic dipole (EMR) and Coulomb quadrupole to
magnetic dipole ratios (CMR).
These can be defined as
$\mbox{EMR} = \Im m \{E_{1+}^{3/2}\} / \Im m \{M_{1+}^{3/2}\}$
and
$\text{CMR} = \Im m \{S_{1+}^{3/2}\} / \Im m \{M_{1+}^{3/2}\}$,
where the pion multipoles, $A^I_{l_{\pi}\pm}$, are characterized through their 
magnetic, electric or longitudinal (scalar) nature, $A=M,E,S$, 
the isospin, $I$, and the pion-nucleon relative angular momentum, $l_{\pi}$, 
whose coupling with the nucleon spin 
is indicated by $\pm$.

Recently, the EMR was determined at squared 4-momentum transfer $Q^2 = 0$ 
with linearly polarized real photons in the reaction 
$ p (\vec \gamma, p) \pi^0 $ \cite{Beck97}.
The result is $ \text{EMR} = (-2.5 \pm 0.2 \pm 0.2) \%$,  
in agreement with the combined partial wave analysis from
$ p (\vec \gamma, p) \pi^0 $  and  $ p (\vec \gamma, \pi^+) n $ 
which allows an isospin decomposition of the multipoles  
\cite{Blanpied97,Hanstein98,Beck00}.

The determination of the longitudinal quadrupole mixing requires 
pion electroproduction experiments.
In the one-photon exchange approximation
the $N(e,e'\pi)N$  cross section can be 
split into photon flux and the virtual photon cross section.
Without target or recoil polarization, the latter
is given by \cite{DT92}
\begin{eqnarray}
\frac{d\sigma_v}{d\Omega_{\pi}} &=& \lambda \cdot
        [      R_T + \epsilon_L R_L 
               + \sqrt{2 \epsilon_L (1+\epsilon)} R_{LT} \cos \Phi 
               \nonumber \\
               & & + \epsilon R_{TT} \cos 2\Phi
               + P_e \sqrt{2 \epsilon_L (1-\epsilon)} R_{LT'} \sin \Phi
        ].
\label{eq:x-sec}
\end{eqnarray}
The structure functions, $R_i$, parameterize the response of the hadronic system
to the various polarization states of the photon field, 
which are described by the transverse and longitudinal
polarization, $\epsilon$  and $\epsilon_L$, respectively,
and by the longitudinal electron polarization, $P_e$.
The ratio $\lambda = |\vec p_\pi^{\,cm}|/k_\gamma^{cm}$  is determined by the
pion cm momentum $\vec p_{\pi}^{\,cm}$  and 
$k_{\gamma}^{cm} = \frac{1}{2} (W-\frac{m_p^2}{W})$, which  
is the photon equivalent energy for the excitation of the target with mass 
$m_p$  to the cm energy $W$.
$\Phi$  denotes the tilt angle between electron scattering 
and reaction plane.

Due to the smallness of the $E_{1+}^{3/2}$  and $S_{1+}^{3/2}$  multipoles 
compared to the dominating $M_{1+}^{3/2}$  amplitude, 
all experimental information so far is based on the extraction of 
$ \Re{e}\{E_{1+}^* M_{1+}\} $  and $ \Re{e}\{S_{1+}^* M_{1+}\} $  
from the $R_{TT}$  and $R_{LT}$  structure functions 
\cite{Beck97,Blanpied97,Siddle71,Alder72,Batzner74,Kalleicher97,Frolov99,Mertz99,Gothe00}.
The interference terms are closely related to the EMR and CMR, respectively.
Due to the unavoidable non delta-resonant contributions
in the structure functions,
it is important to measure the resonant amplitudes 
in several different combinations with the background amplitudes.

This possibility is offered by double polarization observables
\cite{Lourie90,HS98}, 
which furthermore benefit from their insensitivity to experimental
calibration uncertainties.
The general cross section for the
$p (\vec e, e' \vec p\,) \pi^0$  reaction with longitudinally polarized 
electrons and measurement of the recoil proton polarization
is composed of 18 structure functions \cite{DT92,RD89}.
In parallel kinematics, where the proton is detected along the direction
of the momentum transfer, the three components of the proton polarization
take the simple form \cite{HS98} 
(here the notation of ref.\cite{DT92} is used):
\begin{eqnarray}
\sigma_0 P_x &=& \lambda \cdot P_e \cdot \sqrt{2 \epsilon_L (1-\epsilon)} R_{LT'}^t 
\label{eq:P_x_R} \\
\sigma_0 P_y &=& \lambda \cdot \sqrt{2 \epsilon_L (1+\epsilon)} R_{LT}^n
\label{eq:P_y_R} \\
\sigma_0 P_z &=& \lambda \cdot P_e \cdot \sqrt{1-\epsilon^2} R_{TT'}^l.
\label{eq:P_z_R}  
\end{eqnarray}
The proton polarization independent cross section $\sigma_0$  is dominated
by $|M_{1+}|^2$.
The axes are defined by
$ \hat{y} = \vec k_i \times \vec k_f / |\vec k_i \times \vec k_f|$, 
$ \hat{z} = \vec q / |\vec q\,|$  and
$ \hat{x} = \hat{y} \times \hat{z}$,
where $\vec k_i$  and $\vec k_f$  are the momenta of incoming and scattered
electron, respectively.

The structure functions of Eq.\,(\ref{eq:P_x_R})-(\ref{eq:P_z_R}) 
can be decomposed into multipoles of the $\pi^0 p$  final state, 
which are related to the isospin multipoles by
$ A_{l\pm} = A_{l\pm}^{1/2} + \frac{2}{3} A_{l\pm}^{3/2} $.
Displaying only the leading resonance terms with the dominant $M_{1+}$  
amplitude, the polarization components simply read
\begin{eqnarray}
\sigma_0 P_x &=& \lambda \cdot P_e \cdot 
                 c_- \cdot \eta \cdot
                 \Re{e}\{4 S_{1+}^*M_{1+} + n.l.o. \}  
\label{eq:P_x_simple}                                      \\
\sigma_0 P_y &=& 
                 - \lambda \cdot c_+ \cdot \eta \cdot
                 \Im{m}\{4 S_{1+}^*M_{1+} + n.l.o. \}
\label{eq:P_y_simple}                                      \\
\sigma_0 P_z &=& \lambda \cdot P_e \cdot\sqrt{1-\epsilon^2} |M_{1+}|^2 + n.l.o.,
\label{eq:P_z_simple}
\end{eqnarray}
where $\eta=\omega_{cm}/|\vec q_{cm}|$  denotes the ratio of cm energy 
and momentum transfer and
$c_{\pm}=\sqrt{2\epsilon_L(1\pm\epsilon)}$.
As a consequence of the $M_{1+}^{3/2}$  dominance, $\Re{e}\,M_{1+}$ vanishes
very close to the resonance position ($W=1232$\,MeV).
In this case the non-leading order ($n.l.o.$) contributions to $P_x$  are
mainly determined by interferences of imaginary parts of background amplitudes
with $\Im{m}\,M_{1+}$. 
Since the background is predominantly composed of almost purely 
real amplitudes, only small $n.l.o.$  contributions to $P_x$  
are expected.
The situation is different for $P_y$, because here all Born multipoles
contribute, including higher partial waves.
The large $P_y \simeq 40 \%$  that has recently been measured at 
MIT-Bates \cite{Warren98} should, therefore, not a priori be interpreted 
as evidence for non-resonant (imaginary!) background amplitudes in $P_x$.

Eqs.\,(\ref{eq:P_x_simple}) and (\ref{eq:P_z_simple}) show that,
in the $p \pi^0$  channel,
$\widetilde{\mathrm{CMR}} = \Re{e}\{S_{1+}^* M_{1+}\} /|M_{1+}|^2$
can be almost directly determined through either $P_x$  or the 
polarization ratio $P_x/P_z$  \cite{HS98}. 
Furthermore, 
the ratio $R_L/R_T$  of longitudinal
to transverse response is model-independently
accessible without Rosenbluth-separation \cite{HS-LT00}.

At the Mainz Microtron MAMI \cite{Herminghaus90} a
$p(\vec e, e'\vec p\,) \pi^0$  experiment with longitudinally polarized electron 
beam and measurement of the recoil proton polarization has been performed.
The polarized electrons were produced by photoemission from strained GaAsP
crystals using circularly polarized laser light \cite{Aulenbacher97}.
During the experiment the beam helicity was randomly flipped with a frequency 
of 1\,Hz in order to eliminate instrumental asymmetries.
Longitudinal polarization after acceleration was achieved by fine-tuning of
the energy of the microtron to $854.4$\,MeV.
A 5\,cm thick liquid hydrogen target was used. 
Beam currents up to 15\,$\mu$A
with an average polarization of $P_e = 75$\,\% were available.

The scattered electrons were detected at $\theta_e = 32.4^\circ$  
in Spectrometer B of the three spectrometer setup of the A1-collaboration 
\cite{Blomqvist98}.
At $Q^2 = 0.121$\,(GeV/c)$^2$  a range of invariant energies of 
$W = 1200$  to $1260$\, MeV was covered.
The recoil protons were detected in Spectrometer A 
at an angle of $\theta_p = 27^\circ$
with a coincidence time resolution of 1\,ns.
The unobserved $\pi^0$  was identified with a resolution
of $3.8$\,MeV (FWHM) via its missing mass.  
After all cuts, the experimental background was reduced to $0.4$\,\% of the
accepted $\pi^0$  events and thus neglected.

The proton polarization was measured with a focal plane polarimeter 
using inclusive p$-^{12}$C scattering \cite{Pospischil00}.
The detector package of Spectrometer A, consisting of two double
planes of vertical drift chambers for particle tracking and two planes of
scintillators for triggering purposes \cite{Blomqvist98}, was supplemented by
a 7\,cm thick carbon scatterer followed by two double planes of horizontal
drift chambers. 
This setup allowed the eventwise reconstruction of the proton carbon scattering
angles $\Theta_C$  and $\Phi_C$  with a resolution of 2\,mrad.
The proton polarization could be extracted from the azimuthal modulation 
of the cross section
\begin{equation}
\sigma_C = \sigma_{C,0} \left[ 
           1+A_C(P_y^{fp}\cos\Phi_C - P_x^{fp}\sin\Phi_C) \right].
\end{equation}
$\sigma_{C,0}$  denotes the polarization-independent part of the inclusive 
cross section and $A_C$  the analyzing power, which was parameterized according
to \cite{McNaughton85} for $\Theta_C < 18.5^\circ$  and according to
\cite{Pospischil00} for $\Theta_C \geq 18.5^\circ$.

The two polarization components $P_x^{fp}$  and $P_y^{fp}$  are
measured behind the spectrometer's focal plane. 
It is possible to determine
all three components at the electron scattering vertex due to the spin 
precession in the spectrometer and the additional information provided by 
the helicity flip of the electron beam.
The spectrometer's `polarization-optics' is described by a 5-dimensional
spin precession matrix \cite{Pospischil00}.
It was checked through
a series of measurements of the focal plane 
polarization of protons from elastic $p(\vec e, e'\vec p\,)$  scattering  
where the polarization is given by electron kinematics and the 
proton elastic form factors.
From these measurements also the absolute value of $P_e$  was determined,
because in this case the error of $A_C$  cancels out in the determination of
the quantities $P_x/P_e$  and $P_z/P_e$.   

The spin-precession
matrix was used for the extraction of the recoil proton 
polarization in the $p(\vec e, e'\vec p\,)\pi^0$  experiment.
In order to account for the finite acceptance around nominal parallel 
kinematics
($W=1232\,\mbox{MeV}, Q^2=0.121\,\mbox{(GeV/c)}^2, \epsilon=0.718$),
averaged polarizations $\bar P_{x,y,z}^{\rm MAID}$  were generated with
the Mainz Unitary Isobar Model (MAID2000) \cite{Drechsel99} 
for the event population of the experiment.
With the ratios of nominal to averaged polarization,
$\rho_{x,y,z} = P_{x,y,z}^{\rm MAID} / \bar P_{x,y,z}^{\rm MAID}
 = 1.247$, $1.238$  and $0.946$, respectively, 
the experimental polarizations were extrapolated to nominal parallel
kinematics:
$P_{x,y,z} = \rho_{x,y,z} \cdot \bar P_{x,y,z}$.
An additional systematic error is assigned to the polarizations
due to the model uncertainty in $\rho_{x,y,z}$.
It is estimated by a 
$\pm 5\,\%$  variation of the $M_{1+}$  multipole and a  
$\pm 50\,\%$  variation of the other multipoles in MAID.

The results are summarized in Table\,1 along with the statistical and 
systematical errors.
$P_x$  and $P_z$  are given normalized to the beam polarization.
Under the assumptions that the $n.l.o.$  corrections in 
Eqs.\,(\ref{eq:P_x_simple}) and (\ref{eq:P_z_simple})
as well as the 
isospin-1/2 contributions in
$\widetilde{\mathrm{CMR}}$  can be neglected,
the quantities
$ S = \frac{1}{4 \eta c_-} P_x/P_e $  and 
$ R = \frac{\sqrt{1-\epsilon^2}}{4 \eta c_-}
      \frac{P_x/P_e}{P_z/P_e} $ 
of the last two columns can be identified with the CMR.

In contrast to the beam helicity independent $P_y$, 
false systematic asymmetries and systematic errors of the analyzing power
have no impact on $P_x/P_e$  and $P_z/P_e$.
On the other hand, the error caused by drifts of the beam polarization 
only affects $P_x/P_e$  and $P_z/P_e$.
The uncertainty
of the spin tracing affects all three polarization components and
remains the only experimental systematic error in the ratio $R$,
where both the electron beam polarization and the analyzing power drop out.

\begin{figure}
\centerline{\psfig{figure=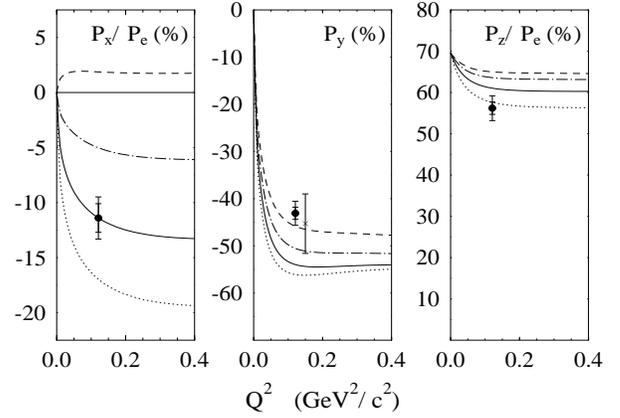,width=8.5cm}}
\caption{ Measured polarization components in comparison with MAID2000
          calculations. The dashed, dot-dashed, full and dotted
          curves correspond to 
          $\text{CMR}=0$, $-3.2$, $-6.4$, $-9.6$\,\%, respectively.
          The MAMI data (full circles) are shown with statistical and 
          systematical error.
          For the Bates $P_y$  (cross) only the statistical error is indicated,
          the value is rescaled in $\epsilon$  and, though measured at the same
          $Q^2$, slightly shifted for clarity.  }
\label{fig:polarizations}
\end{figure}

The results for the three polarization components are shown in Figure
\ref{fig:polarizations} along with MAID2000 calculations.
The curves represent the calculations 
for four values of the CMR: 0, $-3.2$\,\%, $-6.4$\,\% and $-9.6$\,\%.
$P_x$  is most sensitive to the CMR and from this component 
$\text{CMR} = (-6.4 \pm 0.7_{stat} \pm 0.8_{syst})$\,\% is extracted.
The result for the MAID2000 analysis of the ratio $P_x/P_z$  is 
$\text{CMR} = (-6.8 \pm 0.7_{stat} \pm 0.8_{syst})$\,\%.
Both values agree very well and are also close to $S$ and particularly to $R$.
Apparently, the various interference terms in the $n.l.o.$  corrections of 
Eqs.\,(\ref{eq:P_x_simple}) and (\ref{eq:P_z_simple}) 
are small or cancel to a large extent.
From the variation of the multipoles in MAID2000 a model dependence of
the order of 1\,\% absolute of the extracted CMR is estimated.
The 20\,\% discrepancy between MAID and the measured $P_y$  is not expected
to further affect the extracted CMR,
because $P_y$  is more sensitive to interferences other than the CMR.

\begin{figure}
\centerline{\psfig{figure=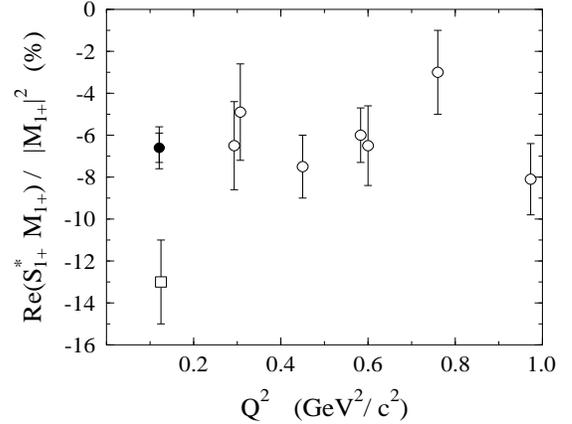,width=8.5cm}}
\caption{ Result for $\Re{e}\{ S_{1+}^* M_{1+} \} / |M_{1+}|^2$  
          as extracted from $P_x/P_e$  of this experiment
          (full circle) with statistical and systematical error,
          compared to unpolarized measurements from
          DESY, NINA, the Bonn synchrotron 
          {\protect \cite{Siddle71,Alder72,Batzner74}}
          (open circles) 
          and ELSA {\protect \cite{Kalleicher97}} 
          (open square),
          where only the statistical errors are indicated.
        }
\label{fig:result}
\end{figure}

In order to compare our result to those from previous
$\pi^0$  electroproduction experiments with unpolarized electrons,
Figure\,\ref{fig:result} shows $\widetilde{\mathrm{CMR}}$ 
instead of CMR.
The MAID2000 analysis of $P_x$  yields 
$\widetilde{\mathrm{CMR}} = (-6.6 \pm 0.7_{stat} \pm 0.8_{syst})\,\%$.
The fact that CMR and $\widetilde{\mathrm{CMR}}$  agree so well demonstrates 
how accurately, at the resonance position, 
the $p \pi^0$  channel yields the CMR
which is defined in the isospin $\frac{3}{2}$  channel.
Our result agrees with older data \cite{Siddle71,Alder72,Batzner74}
and preliminary new Bonn results \cite{Gothe00},
but a recent ELSA result \cite{Kalleicher97} seems to be incompatible
with all other data. 
Whether this has experimental or statistical origin, or points to
an unexpected $\Im{m}S_{0+}$  background contribution
--- which was neglected in \cite{Kalleicher97} --- is still undecided
\cite{HS98a}.

The ratio $R_L/R_T$  of longitudinal to transverse response can be 
obtained in various ways from the so-called reduced polarizations (RPs)
$\chi_{x,y,z}$ \cite{HS-LT00,Kelly99}. 
However, the results vary significantly. 
The smallest value,
$R_L/R_T = (4.7 \pm 0.4_{stat} \pm 0.6_{syst})\,\%$,  
is extracted from the quadratic sum $\chi_x^2+\chi_y^2$  
of the transverse RPs and the largest one,
$R_L/R_T = (12.2^{\,+ 1.7}_{\,-1.6_{stat}}  \,
                        ^{\,+ 2.9}_{\,-2.7_{syst}}) \,\%$,
from $\chi_z$  alone.
Despite the non-linear error propagation in $R_L/R_T(\chi_z)$,
the probability is only a few percent that this discrepancy has purely 
statistical origin.
As a consequence, the consistency relation between the transverse and the 
longitudinal RPs derived in \cite{HS-LT00} seems to be violated.
This presently prohibits a reliable extraction of $R_L/R_T$  but
stresses the importance of a simultaneous measurement of all polarization
components with further improved accuracy.

In summary, we have measured the recoil proton polarization 
in the reaction $p(\vec e, e'\vec p\,)\pi^0$  at the energy of the 
$\Delta(1232)$  resonance.
Due to the spin precession in the magnetic spectrometer all three polarization
components $P_x$, $P_y$  and $P_z$  were simultaneously accessible.
From $P_x$ the Coulomb quadrupole to magnetic dipole ratio was determined 
in the framework of the Mainz Unitary Isobar Model as
$\text{CMR}(Q^2=0.121\,\text{GeV}^2) = (-6.4\pm 0.7_{stat}\pm 0.8_{syst})$\,\%,
which is in good agreement with the result obtained from the polarization ratio
$P_x/P_z$.
There is only a moderate model dependence expected to remain. 
However, the consistency relation among the polarization components seems to be
violated.

The excellent operation of the accelerator and the polarized source 
by K. Aulenbacher, H. Euteneuer, and K.H. Kaiser and their groups
is gratefully acknowledged.
We thank R. Beck for many discussions.
This work was supported by the Deutsche Forschungsgemeinschaft 
(SFB\,443), the Schweizerische Nationalfonds and the
U.S. National Science Foundation.

%
%
\newpage
\begin{table}
\begin{tabular}{|r|c|c|c|c|c|}
\hline
Observable       & $     P_x/P_e$\,(\%) & $     P_y$\,(\%) & $     P_z/P_e$\,(\%)               & $     S$\,(\%)       & $     R$\,(\%)       \\
\hline
Measurement      & --11.4     &  --43.1   &  56.2     &  --5.2    &  --6.4   \\
stat. error      & $\pm$1.3   & $\pm$1.3  & $\pm$1.5  & $\pm$0.6  & $\pm$0.7 \\
\hline
false asymm.     &    --      & $\pm$1.61 &     --    &   --      & --       \\
$\delta P_e$     & $\pm$0.45  &     --    & $\pm$2.25 & $\pm$0.21 & --       \\
$\delta A_C$     &    --      & $\pm$0.87 &     --    &   --      & --       \\
spin precession  & $\pm$1.05  & $\pm$0.45 & $\pm$0.81 & $\pm$0.48 & $\pm$0.62\\
$\delta\rho$     & $\pm$0.9   & $\pm$1.2  & $\pm$1.0  & $\pm$0.41 & $\pm$0.51\\
\hline
total systematic & $\pm$1.4   & $\pm$2.2  & $\pm$2.6  & $\pm$0.7  & $\pm$0.8 \\
\hline
\end{tabular}
\caption{Results for the recoil proton polarization in
         nominal parallel kinematics.
         The quadratic sum of the systematic error contributions yields the 
         total systematic error. 
         The extrapolation to nominal parallel kinematics 
         as well as the ratios $S$  and $R$,
         which approximate the CMR, are explained in the text.}
\end{table}

\input{psfig}
\end{document}